\documentclass[a4paper,11pt]{article}

\usepackage{pos}
\usepackage{lipsum} 
\usepackage{siunitx}
\usepackage{tabularx}

\graphicspath{{./}{figures/}}
\usepackage{subfiles}
\usepackage{subfigure}
\usepackage{hyperref}

\usepackage{lineno}

\title{Searching for High-Energy Neutrino Emission from Seyfert Galaxies in the Northern Sky with IceCube}

\ShortTitle{Neutrino Emission from Seyfert Galaxies}

\author{The IceCube Collaboration \\{\normalsize \normalfont(a complete list of authors can be found at the end of the proceedings)}\\}

\emailAdd{ qinrui.liu@icecube.wisc.edu}

\abstract{

The recent detection of TeV neutrino emission from the nearby active galaxy NGC 1068 by IceCube suggests that AGN could make a sizable contribution to the total high-energy cosmic neutrino flux. The absence of TeV gamma rays from NGC 1068, indicates neutrino production originates in the innermost region of the AGN. Disk-corona models predict a correlation between neutrinos and keV X-rays in Seyfert galaxies, a subclass of AGN to which NGC 1068 belongs. Using 10 years of IceCube through-going track events, we report results from searches for neutrino signals from 27 additional sources in the Northern Sky by studying both the generic single power-law spectral assumption and spectra predicted by the disk-corona model. Our results show excesses of neutrinos associated with two sources, NGC 4151 and CGCG 420-015, at 2.7$\sigma$ significance, and at the same time constrain the collective neutrino emission from our source list. 

\vspace{4mm}
{\bfseries Corresponding authors:}
Theo Glauch$^{1}$, Ali Kheirandish$^{2}$, Tomas Kontrimas$^{1}$, Qinrui Liu$^{3,4 *}$, Hans Niederhausen$^{5}$ \\
{$^{1}$ \itshape Technical University of Munich, TUM School of Natural Sciences, Dept. of Physics}\\
{$^{2}$ \itshape Dept. of Physics \& Astronomy and Nevada Center for Astrophysics, University of Nevada, Las Vegas}\\
{$^{3}$ \itshape Dept. of Physics, Engineering Physics \& Astronomy and Arthur B. McDonald Canadian Astroparticle Physics Research Institute, Queen's University}\\
{$^{4}$ \itshape Perimeter Institute for Theoretical Physics}\\
{$^{5}$ \itshape Dept. of Physics and Astronomy, Michigan State University}\\[4mm]
$^*$ Presenter

\ConferenceLogo{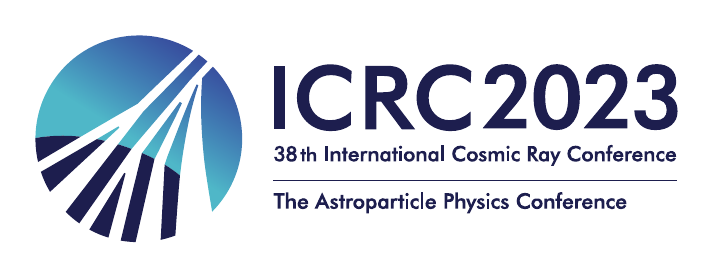}

\FullConference{The 38th International Cosmic Ray Conference (ICRC2023)\\ 26 July -- 3 August, 2023\\ Nagoya, Japan}
}

\begin{document}

\maketitle

\section{Introduction}\label{sec:intro}

The continuous observation of the high-energy neutrino sky by IceCube has recently revealed evidence for particle acceleration in a nearby Seyfert galaxy, NGC 1068~\cite{IceCube:2022der}. This result reinforces the idea that active galactic nuclei (AGN) are cosmic ray (CR) accelerators and make a sizable contribution to the flux of high-energy cosmic neutrinos. With the origin of the rest of the astrophysical neutrino flux unknown, it is well motivated to search for sources similar to NGC 1068.

NGC 1068 was identified as the most significant source so far, with an excess in the energy range of 1.5-15~TeV. The measured neutrino flux from NGC 1068 is much larger than the $\sim$GeV gamma rays measured by {\em Fermi}-LAT~\cite{Fermi-LAT:2019yla,Ballet:2020hze} as well as the upper limits of $\sim$TeV gamma-ray emissions placed by MAGIC and HAWC~\cite{MAGIC:2019fvw,willox2022hawc}. As the interactions of CRs simultaneously produce high-energy neutrinos and gamma rays at the same flux level, the observations indicate that the environments where neutrinos are produced must be opaque to the accompanying gamma rays. The primary candidate is the core of AGN, which can accommodate the efficient production of neutrinos and simultaneously provide an optically thick region where gamma rays are obscured~\cite{Murase:2019vdl}. At the same time, the measurement of the total neutrino flux shows that the flux at medium energies $\left (\sim 30\,\rm{TeV} \right)$ is an order of magnitude greater than that of high energies $\left (\gtrsim 100\,\rm{TeV} \right)$, which implies that sources dominating the medium energies should be opaque to gamma rays in order not to exceed the isotropic gamma-ray background.

\section{Seyfert Galaxies as High-energy Neutrino sources}
In this study, we investigate neutrino emission from the coronae of Seyfert galaxies~\cite{Murase:2019vdl,Kheirandish:2021wkm}. In Seyfert galaxies, accretion dynamics and magnetic dissipation lead to the formation of a hot, highly magnetized, and turbulent corona above the disk~\cite{Miller:1999ix}. The dense environments near the supermassive black hole provides suitable conditions for the interactions of CRs and simultaneous absorption of the accompanying gamma rays. These models, commonly referred to as disk-corona models, can accommodate the excess of neutrino flux at medium energies and the observed flux from NGC 1068~\cite{Murase:2019vdl,Inoue:2019yfs,Kheirandish:2021wkm,Murase:2022feu,Eichmann:2022lxh}. Here, we employ the predicted neutrino flux from the disk-corona model presented in~\cite{Murase:2019vdl,Kheirandish:2021wkm}. In this model, CRs are accelerated stochastically by plasma turbulence in coronae and then interact with gas or radiation in the innermost regions of the AGN to produce neutrinos. AGN coronae are primarily characterized by thermal X-ray emission, making the intrinsic X-ray luminosity $L_X$ the principal parameter in disk-corona models for estimating the neutrino emission. Other model parameters include the CR to thermal pressures that summarizes the CR budget and the turbulence strength. While moderate values of CR to thermal pressure can explain the medium-energy neutrino flux, a higher level of CR pressure is needed to explain the neutrino flux measured in the direction of NGC 1068. This assumption is heavily tied to the measured intrinsic X-ray flux.
For this study, we solely focus on the high CR pressure scenario, given that identification of sources with moderate CR pressure requires next-generation neutrino telescopes. 

Based on the reported intrinsic X-ray flux, this model also finds NGC 1068 as the brightest source in IceCube and suggests that additional sources might be identified if they share similar characteristics with NGC 1068. Here, we conduct analyses focusing on potential neutrino emission from X-ray bright Seyfert galaxies with IceCube muon track events from the Northern Sky (declination~>~-5$^\circ$) for the good pointing power of track events and effective suppression of overwhelming atmospheric muons of up-going events with the Earth acting as a filter.

\section{Analyses}\label{sec:analyses}

\begin{figure}[t!]
\centering
\includegraphics[width=0.65\textwidth]{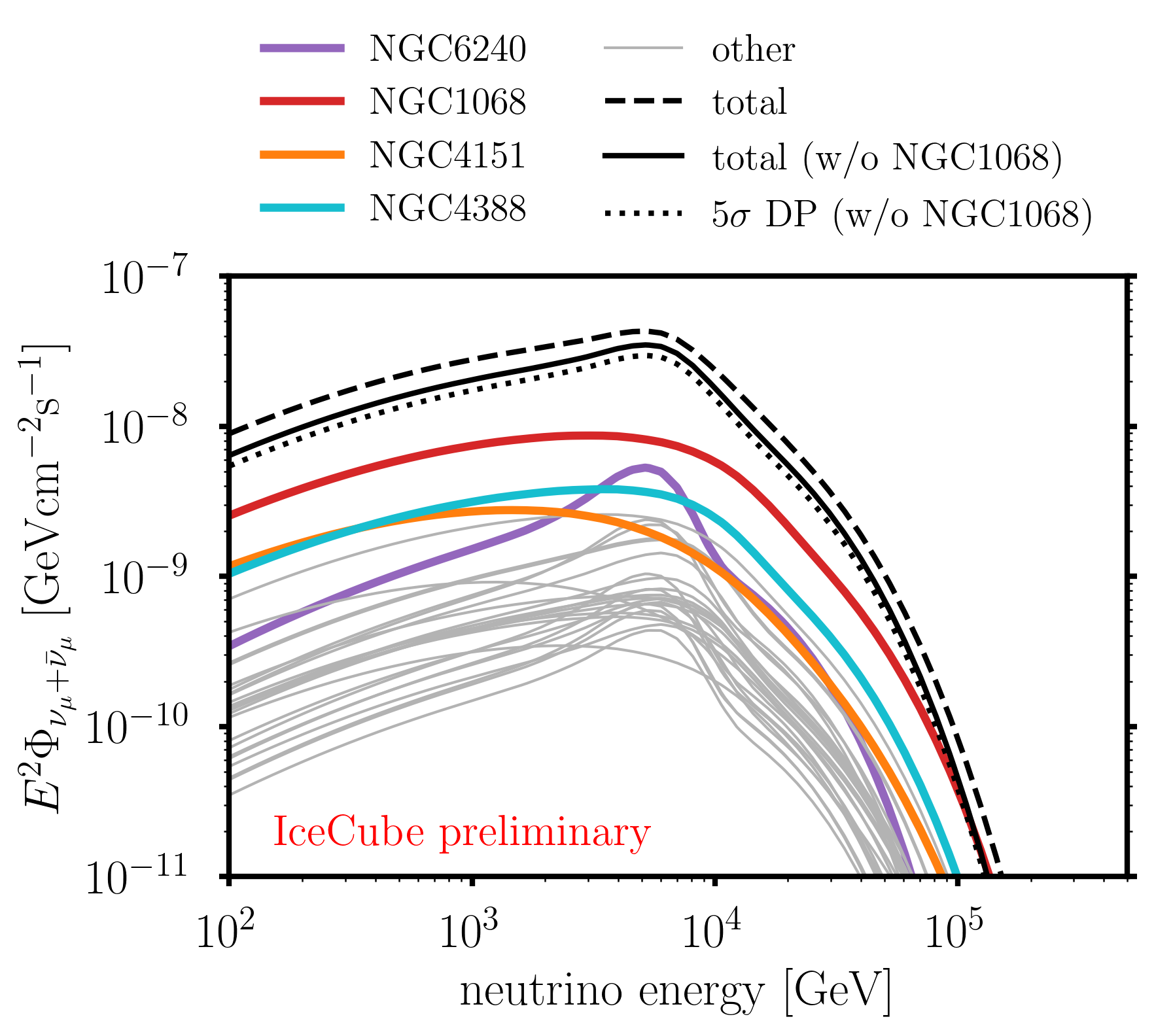}
\caption{The expected flux of each source (thin lines) from the disk-corona model with the top 4 sources, which are likely to be observed by IceCube, are highlighted. The total fluxes excluding or including NGC 1068 are shown, to be compared with the $5\,\sigma$ discovery potentials in both cases.}
\label{fig:model_flux_and_sensitivity}
\end{figure}

Our source selection is based on the BAT AGN Spectroscopic Survey (BASS)~\cite{Ricci:2018eir} which is
an all-sky study of X-ray detected AGN. In the selection, we pick bright Seyfert galaxies in the Northern sky according to their reported intrinsic X-ray fluxes at 2-10 keV as sources with weak X-ray fluxes are not expected to produce detectable neutrino fluxes. NGC 1068 is one of the brightest in this list. 
The selection retains 28 sources in the Northern Sky, including NGC 1068. Considering the knowledge of a strong flux from this source, including NGC 1068 in the list would cause a bias in a search. Therefore, we discuss the exclusion and inclusion of NGC 1068 separately. To be conservative and to take into account the fact that the remaining sources can still give neutrino signals significant enough based on the model expectation, we conclude our results without NGC 1068 and the results including NGC 1068 are shown for completeness. 

In this work, we analyze the $\nu_\mu$ induced muon tracks from the Northern sky. The data sample is processed the same way as in~\cite{IceCube:2022der} which includes new data processing, data calibration, and event reconstruction implemented
that grant us with substantially improved energy reconstructions and point spread function at low to medium energies. In addition to the data used in the previous work, 1.7 years of experimental data was added to the sample. This extension of livetime increases the statistics by $\sim 20\%$ compared to data used in~\cite{IceCube:2022der}. 

We employ the unbinned maximum likelihood ratio method for this work based on the direction, energy proxy, and angular uncertainty of the events in order to discriminate potential neutrino emission from the background composed of the atmospheric and the isotropic astrophysical neutrinos. We perform two types of searches. One is the catalog search, looking for the neutrino emission from each source separately, using power-law and model fluxes, respectively. In addition, we conduct a binomial test to examine the significance of observing excesses of $k$ sources for the two flux hypotheses for our catalog search. The other is the stacking search, where the emission from all selected sources is combined in order to obtain an enhanced signal above the background. In the stacking analysis, only the model flux is tested. 

We apply the improved kernel density estimation (KDE) method presented in~\cite{IceCube:2022der} to these analyses to generate the probability density functions (PDFs). This method improves the modeling of directional distributions of neutrinos significantly compared to the multivariate Gaussian approximation used in previous IceCube analyses. The application of the KDE method depends on the shape of the energy spectrum. For the analyses assuming the disk-corona model, the flux shape varies with $L_X$ and the flux normalization changes with the CR pressure. Other parameters in the calculation are fixed to values fitting the observed flux from NGC 1068 assuming all sources to be intrinsically similar to NGC 1068. Accordingly, we apply KDE to generate the grid of PDFs for the model flux analyses based on $L_X$. As the shape of the flux is determined by the X-ray luminosity, the only free parameter to be fitted in the search is the number of signal $n_s$, which decides the flux normalization.  The expected fluxes of selected sources when setting parameters to ones fitting NGC 1068 are shown in Fig.~\ref{fig:model_flux_and_sensitivity}. The total model fluxes with and without NGC1068 for the stacking search are also shown with comparison to the 5$\sigma$ discovery potential. Even excluding the contribution from NGC 1068, the expected emission exceeds the discovery criterion assuming the optimistic model scenario, i.e., high CR pressure. The analysis performance inspection shows that if the disk-corona model predicts the true flux, modeling the flux correctly gives a notable improvement comparing to fitting the power-law spectrum. The quantity of this improvement is source-dependent. 


As stated above, in addition to the catalog search and stacking search based on the fluxes predicted by the disk-corona model, we also perform a catalog search with the power-law spectrum assumption where the spectral index $\gamma$ is fitted as well as $n_s$. This search has the same procedure as in~\cite{IceCube:2022der} and we continue to use the PDF generation of each spectral index for the power-law flux. This analysis is to complement the search discussed above for possible high-energy events which would be missed due to the cutoff of the model spectrum at high energies and for an intuitive comparison with other work by applying the usual power-law flux assumption.

\begin{table*}[t!]
\small
\centering

\begin{tabular*}{1.\textwidth}{p{2.7cm}p{1.7cm}p{0.6cm}p{0.4cm}p{0.4cm}p{2.5cm}p{2.5cm}p{0.4cm}}

\hline


\phantom{$-$} & spectrum & $n_{\rm exp}$ & $\hat{n}_{\rm s}$ & $\hat{\gamma}$ & $p_{\rm local}$ & $p_{\rm global}$ &$n^{90\%}_{\rm UL}$\\ 

\hline

Stacking Searches & \phantom{$-$} & \phantom{$-$} & \phantom{$-$} & \phantom{$-$} & \phantom{$-$} & \phantom{$-$} & \phantom{$-$} \\\hline
Stacking (excl.) & \textnormal{disk-corona} & 154.4 & 5 & - & $2.4\times 10^{-1}\,(0.7\,\sigma)$ & $2.4\times 10^{-1}\,(0.7\,\sigma)$ & 51.1 \\
Stacking (incl.) $^{(*)}$ &  \textnormal{disk-corona} & 198.9 & 77 & - & $1.1\times 10^{-4}\,(3.7\,\sigma)$ & $-$ & 128.0 \\
\hline
Catalog Search 1 & \phantom{$-$} & \phantom{$-$}  & \phantom{$-$} & \phantom{$-$} & \phantom{$-$} & \phantom{$-$} & \phantom{$-$} \\\hline
CGCG 420-015 & \textnormal{disk-corona} & 3.2   & 31 & - & $2.4\times 10^{-4}\,(3.5\,\sigma)$ & $6.5\times 10^{-3}\,(2.5\,\sigma)$ & 46.4 \\
NGC  4151 & \textnormal{disk-corona} & 13.1   & 23 & - & $6.4\times 10^{-4}\,(3.2\,\sigma)$ & $-$ & 39.5 \\
NGC  1068 $^{(*)}$ & \textnormal{disk-corona} & 44.6 & 48 & - & $3.0\times 10^{-7}\,(5.0\, \sigma)$ & $-$ & 61.4 \\
\hline
Catalog Search 2 & \phantom{$-$} & \phantom{$-$}  & \phantom{$-$} & \phantom{$-$} & \phantom{$-$} & \phantom{$-$} & \phantom{$-$} \\ 
\hline
NGC  4151  & \textnormal{powerlaw} & $-$ & 30 & 2.7 & $6.4\times 10^{-4}\,(3.2\,\sigma)$ & $1.7\times 10^{-2}\,(2.1\,\sigma)$ & 61.4 \\
CGCG 420-015 & \textnormal{powerlaw} & $-$  & 35 & 2.8 & $3.0\times 10^{-3}\,(2.7\,\sigma)$ & $-$ & 62.1 \\
NGC  1068 $^{(*)}$ & \textnormal{powerlaw} & $-$ & 94 & 3.3 & $8.0\times 10^{-8}\,(5.2\,\sigma)$ & $-$ & 94.9 \\
\hline
\end{tabular*}

\caption{Results for the stacking search and selected results from two catalog searches. Best-fitted signal events $\hat{n}_s$, pre-trial and post-trial $p$-values are shown with the post-trial significance. For the model analysis, expected numbers of events ($n_{\rm exp}$) are listed and for the power-law analysis, best-fitted spectral indices $\hat{\gamma}$ are listed. $n^{90\%}_{\rm UL}$ column shows the 90\% confidence level upper limits of the numbers of signal events. Upper limits assuming power-law spectra are given assuming $\gamma=3$. Results marked with $^{(*)}$ are provided for completeness but are not used to compute final significances because evidence for neutrino emission from NGC 1068 was known prior to this work~\citep{IceCube:2022der, IceCube:2019cia}. 
} 
\label{tab:results}
\end{table*}

\section{Results \& Discussion}\label{sec:results}

\begin{figure*}[t!]
    \centering
     \subfigure{%
        \includegraphics[width=0.316\linewidth]{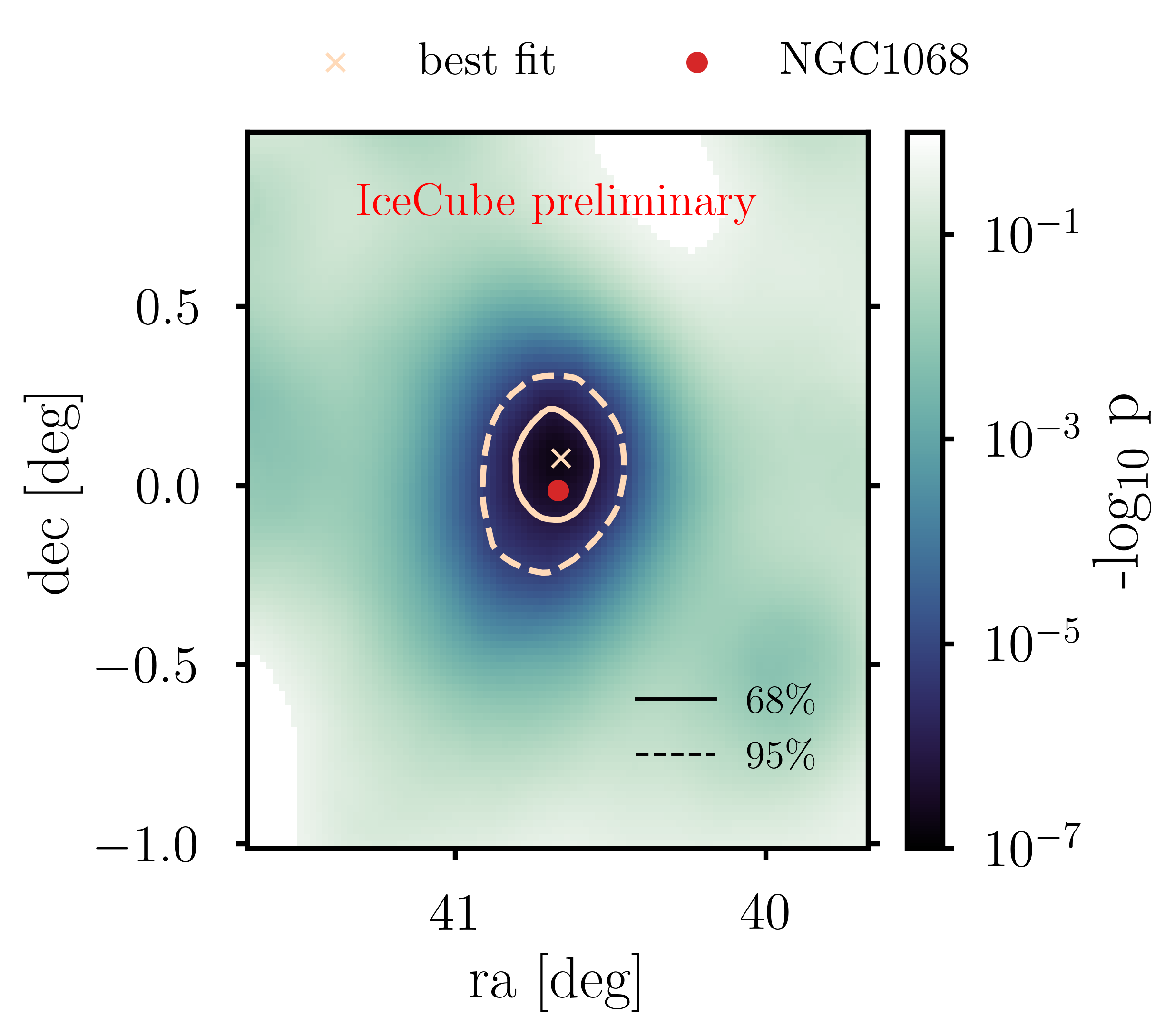}
        \label{fig:gull}}
    \subfigure{
        \includegraphics[width=0.316\linewidth]{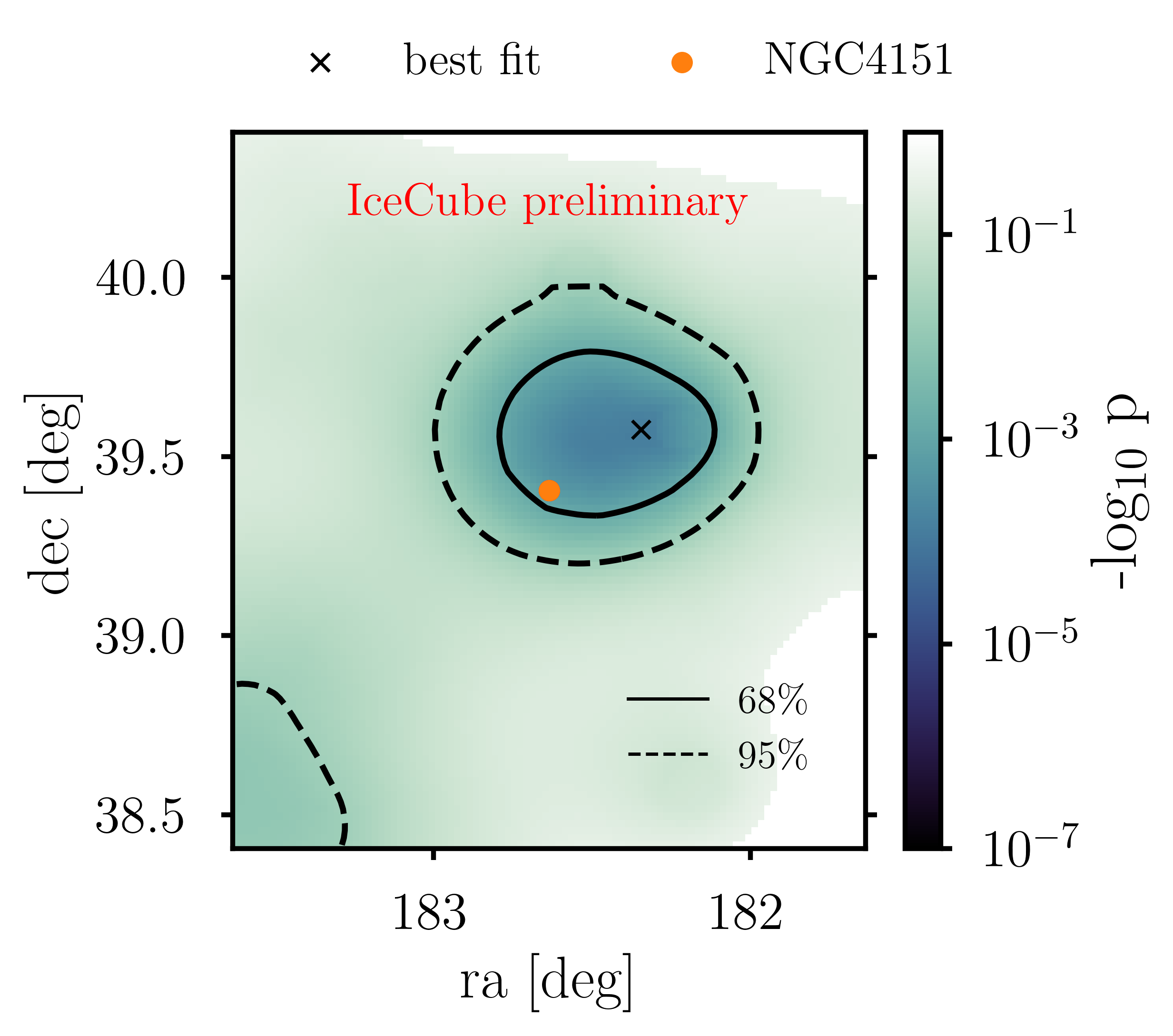}
        }
    \subfigure{
        \includegraphics[width=0.316\linewidth]{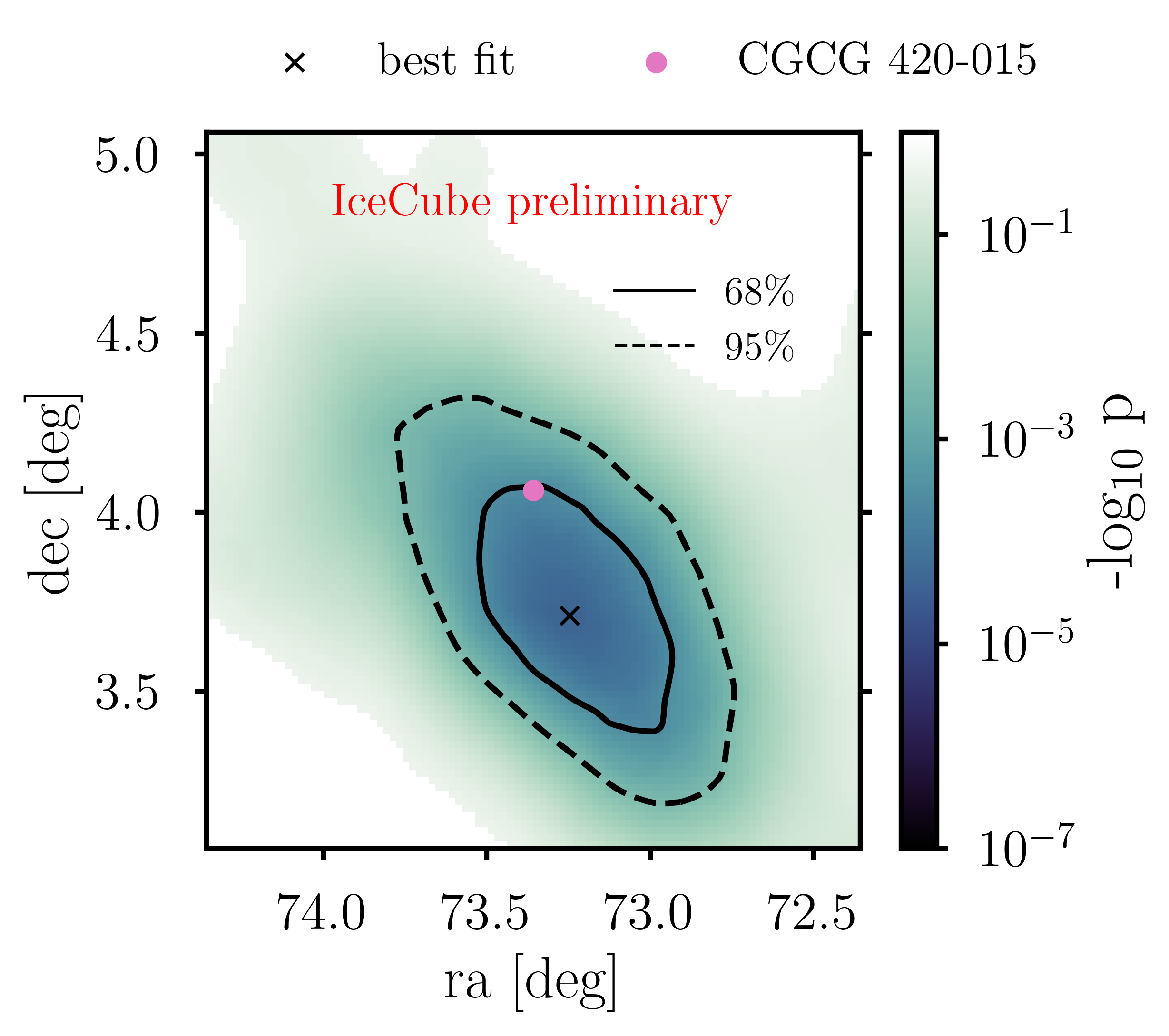}
       }\\     
    \subfigure[]{%
        \includegraphics[width=0.316\linewidth]{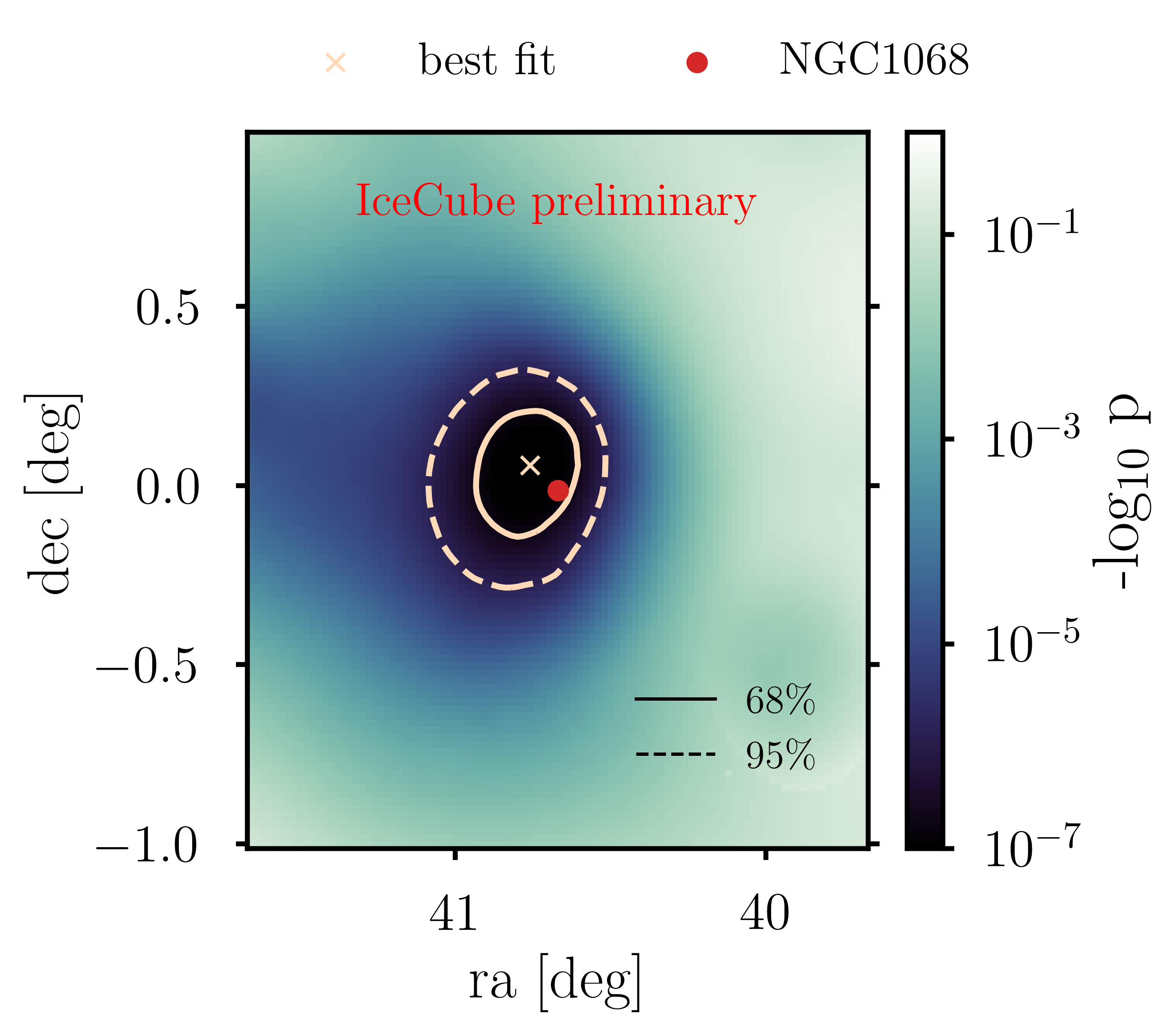}
        }
    \subfigure[]{
        \includegraphics[width=0.316\linewidth]{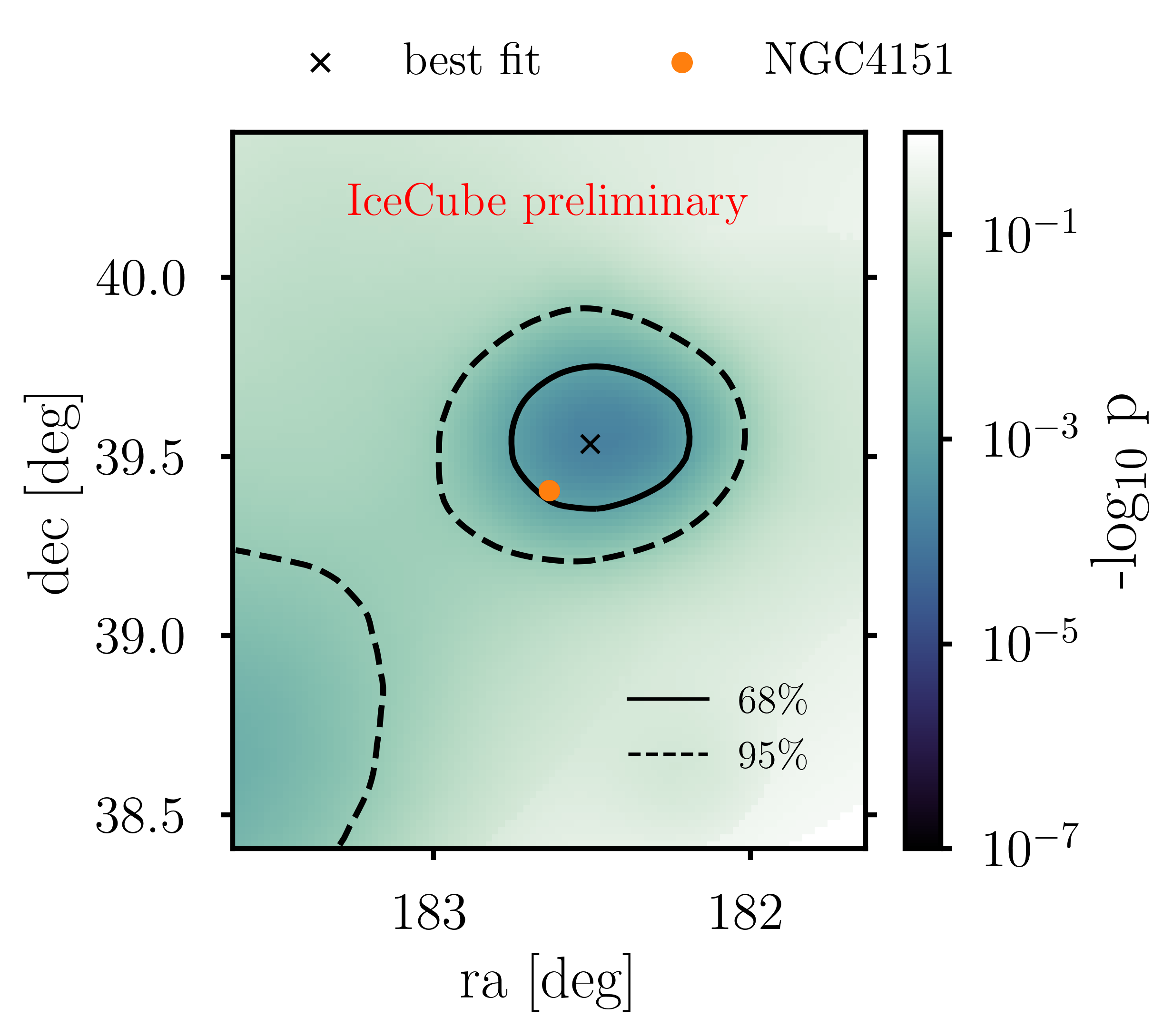}
        }
    \subfigure[]{
        \includegraphics[width=0.316\linewidth]{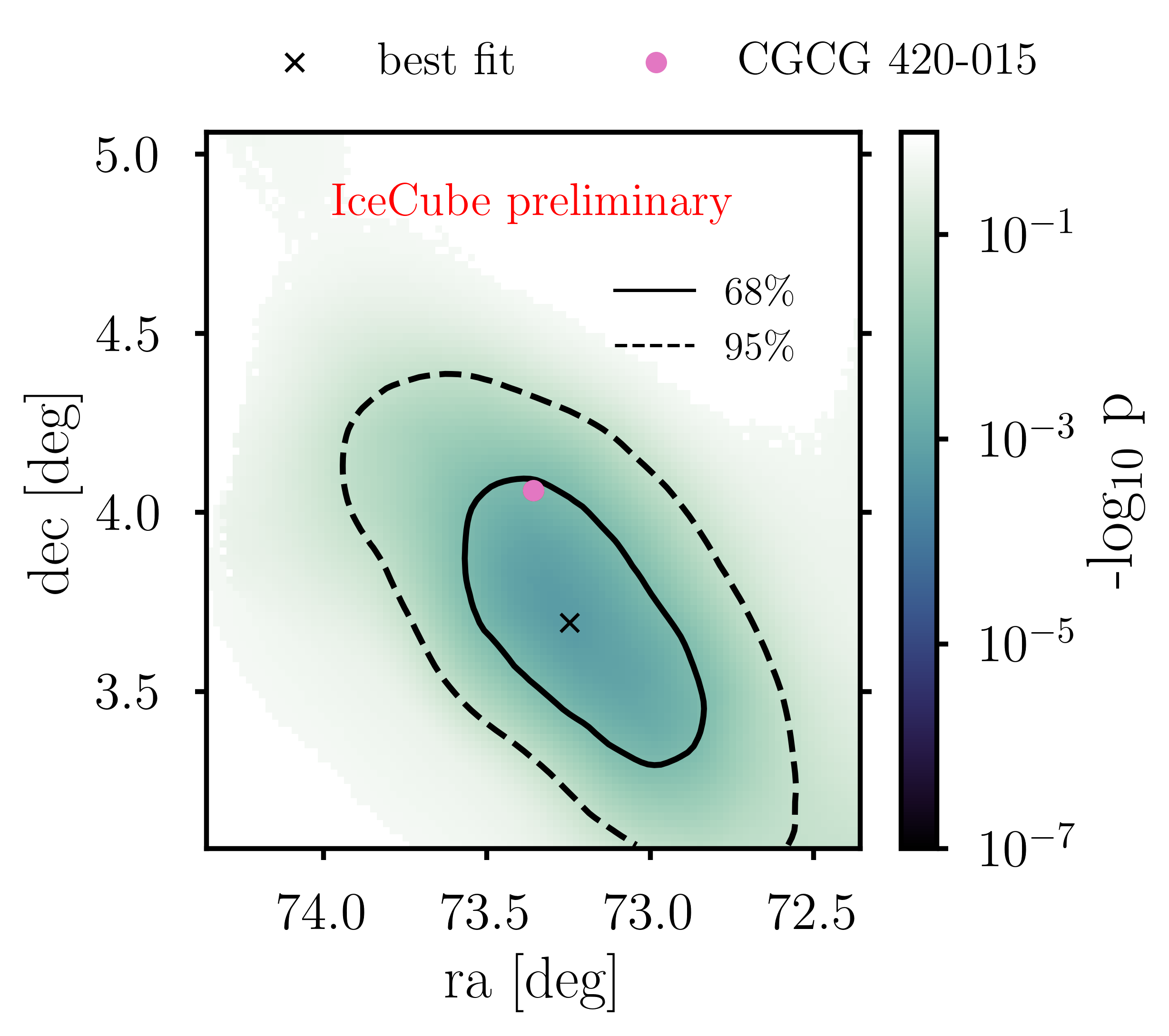}
        }
    \caption{Local pre-trial $p$-value maps around the top sources NGC 1068, NGC 4151 and CGCG 420-015 with the the model fit (top) and the power-law fit (bottom). Colored points show the locations of sources and crosses show the best-fit locations. Contours correspond to 68\% (solid) and 95\% (dashed) confidence regions. }
    \label{fig:p_scan}
\end{figure*}

\begin{figure*}[t!]
\centering
\includegraphics[width=0.68\textwidth]{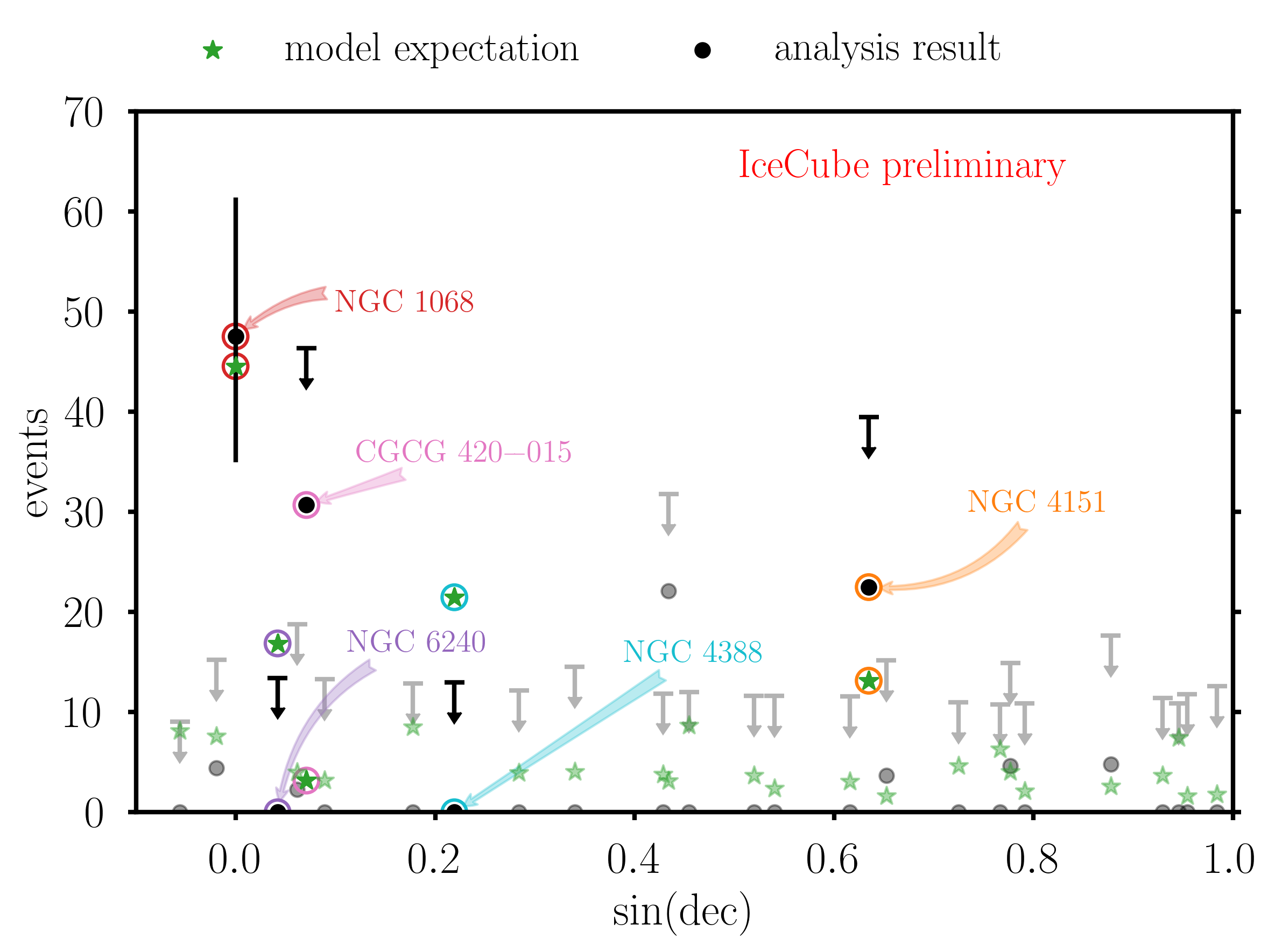}
\caption{Expected numbers of events (green stars) from the model and the best-fitted numbers of signal events (black circles) for individual sources. Down arrows show the 90\% upper limits. The top 4 sources predicted by the model listed in Fig.~\ref{fig:model_flux_and_sensitivity} which include NGC 1068 and NGC 4151 are highlighted, together with the most significant source in the search assuming the model flux, i.e. CGCG 420-015. }
\label{fig:results_summary}
\end{figure*}

The results for the top sources in the two catalog searches and the stacking search are summarized in Table~\ref{tab:results}. In addition to NGC 1068, we find that excesses of neutrino emission could be associated with two other sources: CGCG 420-015 and NGC 4151. CGCG 420-015 is the most significant in the search based on the disk-corona model flux assumption with a $2.5\sigma$ post-trial significance while NGC 4151 stands out in the search based on the power-law spectrum assumption with a $2.1\sigma$ post-trial significance. The significance of NGC 1068 increases owing to the increase of the statistics of the data. Fig.~\ref{fig:p_scan} shows the $p$-value scans in the regions around the top sources under our two flux assumptions. For all selected sources, Fig.~\ref{fig:results_summary} displays event numbers of the expectations as well as the measurement with the 90\% confidence level upper limits. The binomial test results in a post-trial 2.7$\sigma$ excess from CGCG 420-015 and NGC 4151 when we exclude NGC 1068 and the significance grows to $4\sigma$ including NGC 1068. There is no significant excess found in the stacking search  with a $p$-value=0.24 without including the contribution from NGC 1068, and the best-fit event number is much below the expectation. The results, on one hand, demonstrate the feasibility of identifying sources similar to NGC 1068 in the catalog searches and the binomial test. On the other hand, the absence of a strong signal in the stacking search implies the model parameters suited to explain the observed neutrino flux from NGC 1068 are unlikely to be shared with most sources in the selected list. 

The first implication of the results is that the CR pressure, which sets the normalization of CRs at the source, is lower than what is fitted for NGC 1068 for most sources. As discussed in~\cite{Kheirandish:2021wkm}, more moderate neutrino emission scenarios are beyond the detectability of current neutrino telescopes and the identification of those sources is more feasible with the next-generation detectors. 

Meanwhile, the selection of bright Seyfert galaxies and the calculation of the expected neutrino flux in the disk-corona model highly depend on the reported intrinsic X-ray flux by BASS, which introduces the primary uncertainty in the analysis as precise estimation of the intrinsic luminosity is challenging for Compton thick sources. Regardless that the BASS catalog offers the most comprehensive survey of non-jetted AGN, more accurate measurement is usually accomplished by targeting instruments such as \emph{NuSTAR}. It is worth mentioning that the higher intrinsic flux from NGC 1068 reported in ~\cite{Marinucci:2015fqo} would indicate lower CR pressure, which would decrease the expected emission from the other sources in the catalog.

\section{Summary}\label{sec3}

In this study, we searched for high-energy neutrino emission from X-ray bright Seyfert galaxies
in the Northern Hemisphere. We incorporate the disk-corona model to perform a catalog search and a stacking
search on our selected sources where the generic power-law spectrum assumption is also applied for a catalog
search. As there is no significant excess of neutrino events observed in the stacking search, we can constrain
the collective neutrino emission from those X-ray bright Seyfert galaxies in Northern sky. However,
our results hint neutrino emission from two sources, i.e. NGC 4151 and CGCG 420-015 in addition to NGC 1068. Our results might implicate the existence of sources similar to NGC 1068 whose neutrino emission can possibly be
explained by the disk-corona model. Nevertheless, the absence of a significant correlation in the stacking search
and most individual sources implies that the features of NGC 1068 leading to the strong neutrino emission are
not commonly shared with other X-ray bright Seyfert galaxies. The expectation of neutrino emission relies considerably on the details of the modeling within the picture of the disk-corona model and more comprehensive multi-wavelength observations will provide further insight on the characteristics of the potential sources which will benefit the modeling significantly. 

IceCube-Gen2, the next-generation of the IceCube detector~\cite{IceCube-Gen2:2020qha}, will be 8 times larger in volume with an expected $\sim$5 times increase of the muon track effective area. The sensitivity to $\nu_\mu$ fluxes is expected to rise similarly. This improvement is expected to provide promising prospects for enhancement of the excess from the interesting sources and potential of finding more sources in the future, including ones expected to have moderate neutrino emission. Considering the fact that the majority of the bright Seyfert galaxies reside in the Southern Sky, the improved sensitivity in this region recently achieved by the technical progress in track events selection by IceCube~\cite{IceCube:2021ctg} provides an opportunity to identify more sources. A similar study focusing on the Southern Sky X-ray bright Seyfert galaxies using this selection is presented in~\cite{IceCube:2023southseyfert}. In the upcoming years, detectors instrumented in the Northern Hemisphere will boost the identification of sources in the Southern Sky, complementing the detection prospect in the Northern Sky.

\bibliographystyle{ICRC}
\bibliography{references}

%

\clearpage

\section*{Full Author List: IceCube Collaboration}

\scriptsize
\noindent
R. Abbasi$^{17}$,
M. Ackermann$^{63}$,
J. Adams$^{18}$,
S. K. Agarwalla$^{40,\: 64}$,
J. A. Aguilar$^{12}$,
M. Ahlers$^{22}$,
J.M. Alameddine$^{23}$,
N. M. Amin$^{44}$,
K. Andeen$^{42}$,
G. Anton$^{26}$
% ...
\\
\\
% ...
$^{12}$ Universit{\'e} Libre de Bruxelles, Science Faculty CP230, B-1050 Brussels, Belgium \\
% ...
$^{17}$ Department of Physics, Loyola University Chicago, Chicago, IL 60660, USA \\
$^{18}$ Dept. of Physics and Astronomy, University of Canterbury, Private Bag 4800, Christchurch, New Zealand \\
% ...
$^{22}$ Niels Bohr Institute, University of Copenhagen, DK-2100 Copenhagen, Denmark \\
$^{23}$ Dept. of Physics, TU Dortmund University, D-44221 Dortmund, Germany \\
% ...
$^{26}$ Erlangen Centre for Astroparticle Physics, Friedrich-Alexander-Universit{\"a}t Erlangen-N{\"u}rnberg, D-91058 Erlangen, Germany \\
% ...
$^{40}$ Dept. of Physics and Wisconsin IceCube Particle Astrophysics Center, University of Wisconsin{\textendash}Madison, Madison, WI 53706, USA \\
% ...
$^{42}$ Department of Physics, Marquette University, Milwaukee, WI, 53201, USA \\
% ...
$^{44}$ Bartol Research Institute and Dept. of Physics and Astronomy, University of Delaware, Newark, DE 19716, USA \\
% ...
$^{63}$ Deutsches Elektronen-Synchrotron DESY, Platanenallee 6, 15738 Zeuthen, Germany  \\
$^{64}$ Institute of Physics, Sachivalaya Marg, Sainik School Post, Bhubaneswar 751005, India \\
% ...

\subsection*{Acknowledgements}

\noindent
The authors gratefully acknowledge the support from the following agencies and institutions:
USA {\textendash} U.S. National Science Foundation-Office of Polar Programs,
U.S. National Science Foundation-Physics Division,
U.S. National Science Foundation-EPSCoR,
Wisconsin Alumni Research Foundation,
Center for High Throughput Computing (CHTC) at the University of Wisconsin{\textendash}Madison,
Open Science Grid (OSG),
Advanced Cyberinfrastructure Coordination Ecosystem: Services {\&} Support (ACCESS),
Frontera computing project at the Texas Advanced Computing Center,
U.S. Department of Energy-National Energy Research Scientific Computing Center,
Particle astrophysics research computing center at the University of Maryland,
Institute for Cyber-Enabled Research at Michigan State University,
and Astroparticle physics computational facility at Marquette University;
Belgium {\textendash} Funds for Scientific Research (FRS-FNRS and FWO),
FWO Odysseus and Big Science programmes,
and Belgian Federal Science Policy Office (Belspo);
Germany {\textendash} Bundesministerium f{\"u}r Bildung und Forschung (BMBF),
Deutsche Forschungsgemeinschaft (DFG),
Helmholtz Alliance for Astroparticle Physics (HAP),
Initiative and Networking Fund of the Helmholtz Association,
Deutsches Elektronen Synchrotron (DESY),
and High Performance Computing cluster of the RWTH Aachen;
Sweden {\textendash} Swedish Research Council,
Swedish Polar Research Secretariat,
Swedish National Infrastructure for Computing (SNIC),
and Knut and Alice Wallenberg Foundation;
European Union {\textendash} EGI Advanced Computing for research;
Australia {\textendash} Australian Research Council;
Canada {\textendash} Natural Sciences and Engineering Research Council of Canada,
Calcul Qu{\'e}bec, Compute Ontario, Canada Foundation for Innovation, WestGrid, and Compute Canada;
Denmark {\textendash} Villum Fonden, Carlsberg Foundation, and European Commission;
New Zealand {\textendash} Marsden Fund;
Japan {\textendash} Japan Society for Promotion of Science (JSPS)
and Institute for Global Prominent Research (IGPR) of Chiba University;
Korea {\textendash} National Research Foundation of Korea (NRF);
Switzerland {\textendash} Swiss National Science Foundation (SNSF);
United Kingdom {\textendash} Department of Physics, University of Oxford.

\end{document}